\documentstyle[preprint,prl,aps]{revtex}

\tightenlines

\input epsf

\title{Running Coupling in the SU(2) Lattice Gauge Theory}

\author{
O.~Borisenko $^1$,
M.~Gorenstein $^{1,2,3}$ and A.~Kostyuk $^1$
}

\address{
$^1$ Bogolyubov Institute for Theoretical Physics,
Kiev, Ukraine\\
$^2$ School of Physics and Astronomy,
Tel Aviv University, Tel Aviv, Israel\\
$^3$ Institute for Theoretical Physics, Goethe University,
Frankfurt, Germany
}

\draft

\begin{document}

\maketitle

\begin{abstract}

Scenario according to which the SU(2)-gluodynamics
is a theory with a nontrivial fixed point is analyzed from the point
of view of the modern  Monte-Carlo (MC) lattice data.
It is found that an {\it assumption} of the first order fixed point
$g=g_f$
of the beta function $\beta_f(g)$ has no contradictions with
existing MC lattice data. The beta function parameters
are found from
the requirement of constant values for critical temperature
$T_c/\Lambda_L^{FP}$ and string tension
$\sqrt{\sigma}/\Lambda_L^{FP}$ in MC lattice calculations at $4/g^{2}
\geq 2.30$.

\end{abstract}


\vskip 0.5cm

The Monte--Carlo (MC) lattice simulations
are one of the main sources of the non-perturbative
results in the gauge field theories.
For SU(N) gluodynamics on
lattices of size $N_{\tau}\times N_{\sigma}^3$
MC results are the dimensionless functions
of a bare coupling constant $\beta = 2N/g^2$.
The transformation of these functions to physical quantities
are done by multiplying them with a lattice spacing $a$
in the corresponding powers.
The length scale $L$ ($V=L^3$ is the system volume)
and the temperature $T$ are given by
\begin{equation}\label{a}
L~=~N_{\sigma}~a~,~~~~  T~=~(N_{\tau}~a)^{-1}.
\end{equation}
The lattices with $N_{\sigma}>> N_{\tau}$
correspond to the finite temperature models, whereas those with
$N_{\sigma}\approx N_{\tau}$ are identified
with zero temperature limit.
Requirement $N_{\tau}>>1$ is necessary
to avoid lattice artifacts.

The best studied quantities in the
MC simulations of pure $SU(N)$ gauge theories are
the dimensionless string tension $(\sqrt{\sigma}a)_{MC}$
and the critical coupling $\beta_c^{MC}$
of the deconfinement phase transition.
The values of $\beta_c^{MC}$ were found
for the finite lattices and
the extrapolation to spatially infinite volume (`thermodynamic limit')
$N_{\sigma}\rightarrow \infty$
has been done (see Ref.~\cite{gc} and references therein).
In what follows we discuss $SU(2)$ gluodynamics.
The MC values of the critical couplings
$\beta_{c}^{MC}$ for different $N_{\tau}$ and of
$(\sqrt{\sigma}a)_{MC}$ for different $\beta$
are presented in Table I and Table II, respectively.
The data are taken from Ref.~\cite{gc}.
To define their physical values one needs a connection
between the lattice spacing $a$ and the bare coupling constant $g$.
Such a connection is given in terms of the beta function
$\beta_f(g)$ through the equation:
\begin{equation}\label{beta}
\beta_{f} (g)~ = ~-~ a \frac{d g}{d a}~.
\end{equation}
The conventional perturbation theory gives the following expansion
of the beta function
\begin{equation}\label{betaaf}
\beta_{f}^{AF} (g)~ =~ -b_0 g^3-b_1 g^5 +O(g^7)~, ~~~~~~
b_0=\frac{11 N}{48 \pi^{2}}~,~~
b_1=\frac{34}{3} \left( \frac{N}{16 \pi^2} \right)^{2}~.
\end{equation}
$N$ refers to the group $SU(N)$.
Differential equation (\ref{beta}) with $\beta_{f}^{AF}(g)$
(\ref{betaaf}) leads to
\begin{equation}\label{agaf}
a\Lambda_L^{AF}~ \cong ~ \exp\left(-\frac{1}{2b_0 g^2}\right)~
(b_0 g^2)^{-b_1/2 b_0^2}~\equiv~R(g^2)~
\end{equation}
where $\Lambda_L^{AF}$ is an integration constant
of Eq.~(\ref{beta}). Eq.~(\ref{agaf}) is known as the asymptotic freedom
(AF) relation.

Using Eqs.~(\ref{a}) and (\ref{agaf}) one can calculate
\begin{equation}\label{Tcaf}
T_c/\Lambda_L^{AF}~ \equiv~
\frac{1}{N_{\tau}a_c\Lambda_L^{AF}}~
=~\frac{1}{N_{\tau} R(g_c^2)}~
\end{equation}
and
\begin{equation}\label{sigmaaf}
\sqrt{\sigma}/\Lambda_L^{AF}~ \equiv~
\frac{(\sqrt{\sigma}a)_{MC}}{a\Lambda_L^{AF}}
~=~
\frac{(\sqrt{\sigma}a)_{MC}}{R(g^2)}~.
\end{equation}
The values of
$T_c/\Lambda_L^{AF}$ (see also Ref.~\cite{gc})
at different $N_{\tau}$
are presented in our Table I
and of $\sqrt{\sigma}/\Lambda_L^{AF}$
for different couplings $\beta$ in Table II.
One observes a rather strong dependence
of $T_c/\Lambda_L^{AF}$ on $N_{\tau}$
and $\sqrt{\sigma}/\Lambda_L^{AF}$ on $\beta$.
It means that
the perturbative AF relation (\ref{agaf})
does not work even on the largest available lattices.
This fact is known as an absence of the asymptotic scaling.
In contrast to the problem with  an asymptotic scaling
the scaling has been observed for the
ratios of different physical quantities
calculated from the lattice expectation values. MC data from
Tables I and II give for different lattices
almost a constant ratio
\cite{gc}:
\begin{equation}\label{ratio}
\left(\frac{T_c}{\sqrt{\sigma}}\right)_{MC}~=~0.69~\pm~0.02
\end{equation}
if MC values for
$T_c/\Lambda_L^{AF}$ (\ref{Tcaf})
and $\sqrt{\sigma}/\Lambda_L^{AF}$
(\ref{sigmaaf}) are calculated at equal coupling constants
$\beta$ in the region  $\beta \geq 2.30$.
It suggests a possibility of the universal asymptotic scaling
violation:
it has been proposed in Ref.~\cite{eng95}
that a deviation from the asymptotic scaling
can be described by a universal `non-perturbative' (NP) beta function,
i.e., $\beta_{f}^{NP}(g)$ is the same for all lattice observables
and it does not depend on the lattice size if $N_{\sigma}$
and $N_{\tau}$ are not too small.

The following ansatz was suggested
\cite{eng95}:
\begin{equation}\label{ansatz}
a\Lambda_L^{NP} ~=~\lambda (g^2) R(g^2)~,
\end{equation}
where $R(g^2)$ is given by Eq.~(\ref{agaf}) and
$\lambda (g^2)$ is thought to describe a deviation from
the perturbative behaviour.
The equation (\ref{agaf}) has been expected
at $g \rightarrow 0$ so that
an additional constraint, $\lambda (0)=1$,
has been assumed.
The values of $T_c/\Lambda_L^{NP}$ and
$\sqrt{\sigma}/\Lambda_L^{NP}$
can be calculated then as
\begin{equation}\label{TcL}
T_c/\Lambda_L^{NP}~ =~
\frac{1}{N_{\tau}~\lambda(g_c^2) R(g_c^2)}~,
\end{equation}
\begin{equation}\label{sigmaL}
\sqrt{\sigma}/\Lambda_L^{NP}
~=~
\frac{(\sqrt{\sigma}a)_{MC}}{\lambda(g^2) R(g^2)}~.
\end{equation}
A simple formula
for the function $\lambda (g^2)$ was suggested \cite{eng95}:
\begin{equation}\label{lambda}
\lambda (g^{2})~ =~ \exp \left(
\frac{c_{3} g^{6}}{2 b_{0}^{2}}
\right)~.
\end{equation}
Parameter $c_3$ in Eq.~(\ref{lambda}) and a new one,
$T^*_c/\Lambda_L^{NP}= const$,
were considered as free parameters
and determined from fitting the
MC values of
$T_c/\Lambda_L^{NP}$ (\ref{TcL})
at different $N_{\tau}$
to the constant value
$T^*_c/\Lambda_L^{NP}$.
This procedure gives:
\begin{equation}\label{param}
T^*_c/\Lambda_L^{NP} ~=~ 21.45(14)~,~~~~
c_{3}~=~5.529(63)\cdot10^{-4}~.
\end{equation}
The numerical values
of $T_c/\Lambda_L^{NP}$ (\ref{TcL})
are presented in our Table I.
In comparison to $T_c/\Lambda_L^{AF}$
one observes much weaker $N_{\tau}$ dependence of
$T_c/\Lambda_L^{NP}$ (\ref{TcL}). They become now close
to the constant value $T_c^*/\Lambda_L^{NP}$ (\ref{param}).
Due to Eq.~(\ref{ratio}) the constancy of
$T_c/\Lambda_L^{NP}$ (\ref{TcL})
guarantees an approximate constancy of the physical string tension
$\sqrt{\sigma}/\Lambda_L^{NP}$ (\ref{sigmaL})
with the average value
$\sqrt{\sigma ^*}/\Lambda_L^{NP}=31.56$
in the region of coupling
constant
$\beta = 2.3\div2.8$ (see Table II).

In spite of the evident phenomenological success
of the above procedure of Ref.~\cite{eng95}
the crucial question regarding the validity of the
perturbative AF relation (\ref{agaf}) at $g\rightarrow 0$
is not solved and remains just a postulate.
Do the existing MC data rule out any other possibility?
To answer this question we reanalyze the same MC data
using the same strategy as in Ref.~\cite{eng95}.
A principal difference of our analysis is
that we do not assume the AF
relation (\ref{agaf}) between $g$ and $a$
at $g\rightarrow 0$. Instead of this standard approach we
check a quite different scenario
with the fixed point (FP) $g_f$ of the beta function.
Note that the so-called FP field theory models were considered a
long time ago \cite{fp}.
This theoretical possibility has been also discussed in Ref.~\cite{fp1}.
It was demonstrated in Ref. \cite{fp2} that the precise data on deep
inelastic scattering do not eliminate the FP model and other
tests would be necessary to distinguish between AF and
FP QCD.

Let us {\it assume}
that the beta function of the SU(2)-gluodynamics
has a zero of the first order
at some FP $g_f$. We thus have in the vicinity of this point
\begin{equation}\label{betafp}
\beta_{f}^{FP}(g)~=~-b~(g~-~g_f)~.
\end{equation}
\noindent
From Eqs.~(\ref{beta}) and (\ref{betafp}) we find then
\begin{equation}\label{afp}
a~\Lambda_L^{FP}~=~(g~-~g_f)^{1/b}~,
\end{equation}
with $\Lambda_L^{FP}$ being an
arbitrary integration constant
of differential equation (\ref{beta}).
It follows then for the critical temperature
\begin{equation}\label{Tcfp}
T_c/\Lambda_L^{FP}~
=~\frac{1}{N_{\tau}(g_c~-~g_f)^{1/b}}~
\end{equation}
and for the string tension
\begin{equation}\label{sqsifp}
\sqrt{\sigma}/\Lambda_L^{FP}~=~
\frac{(\sqrt{\sigma}a)_{MC}}{(g~-~g_f)^{1/b}}~.
\end{equation}
Our requirement
similar to that of Ref.~\cite{eng95}
is to fit
$T_c/\Lambda_L^{FP}$ (\ref{Tcfp}) for
different $N_{\tau}$
to a constant:
\begin{equation}\label{CT}
T^*_c/\Lambda_L^{FP}~\equiv~C_{T}~=~const~,
\end{equation}
where numerical value of $C_{T}$ is a priori
unknown. This requirement
leads to the following expression for the critical coupling
\begin{equation}\label{betac}
g_c~ = ~(C_{T} N_{\tau})^{-b}~ +~ g_{f}~, ~~~
\beta_c~
=~ \frac{4}{[ (C_{T} N_{\tau})^{-b} + g_{f} ]^{2}}~.
\end{equation}
In addition we require the constancy of $\sqrt{\sigma}/\Lambda_L^{FP}$
(\ref{sqsifp}) for different $\beta$. It introduces another
unknown constant
\begin{equation}\label{csigma}
\sqrt{\sigma^*}/\Lambda_L^{FP}~\equiv~C_{\sigma}~=~const
\end{equation}
and leads to the model equation for $\sqrt{\sigma}a$
as a function of $\beta$:
\begin{equation}\label{sigmamc}
\sqrt{\sigma}a~ =~ C_{\sigma} (g~ -~ g_{f})^{1/b}~
~\equiv C_{\sigma} (2/\sqrt{\beta}~ -~ g_{f})^{1/b}~.
\end{equation}

Our fitting procedure for
finding beta function
parameters $g_{f}$ and $b$ as well as the
constants $C_{T}$ (\ref{CT}) and $C_{\sigma}$ (\ref{csigma})
is to minimize $\chi ^2$ defined as
\begin{equation}\label{chi2}
\chi^2~=~ \sum_{N_{\tau}}
\frac{
\left[ (\beta_c~ -
~ \beta_c^{MC}) \right]^{2}
}
{\left[ \Delta \beta_c^{MC} \right]^2}
~+~\sum_{\beta_i}
\frac{
\left[ \sqrt{\sigma}a~-~
(\sqrt{\sigma}a)_{MC}  \right]^2
}
{
\left[ \Delta (\sqrt{\sigma}a)_{MC} \right]^{2}}~,
\end{equation}
where
$\beta_c$ and
$\sqrt{\sigma}a$ are given by model equations (\ref{betac})
and (\ref{sigmamc}),
$\Delta \beta_c^{MC}$ and
$\Delta (\sqrt{\sigma}a)_{MC}$ stand for uncertainties of
the corresponding MC data.
We assume that deviations from the formula (\ref{afp})
are negligible for
$\beta \geq 2.30$
and use available
MC data from Tables I and II satisfying this
criterion: in the first sum of
Eq.~(\ref{chi2}) we include MC
points $\beta_c^{MC}$ from Table I for
$N_{\tau}=5,~6,~8,~16$ and
the second sum is carried over
$\beta_i=2.3,~2.4,~2.5,~2.6,~2.7,~2.85$ with corresponding MC
values of $(\sqrt{\sigma}a)_{MC}$ from Table II.

The $\chi^{2}$ reaches its minimum $\chi^{2}_{min}=2.15$ at
\begin{equation}\label{min}
g_f~=~ 0.563~,~~~ b~=~0.111~,~~~ C_{T}~=~3.15~,~~~ C_{\sigma}~=~4.59~.
\end{equation}

For the set of parameters (\ref{min}) the
values of $T_c/\Lambda_L^{FP}$
(\ref{Tcfp}) and $\sqrt{\sigma}/\Lambda_L^{FP}$ are presented in
Tables I and II. They are constant within
the errors\footnote{ The errors are induced solely by the uncertainties
of the MC data for $\beta_c^{MC}$ and $(\sqrt{\sigma}a)_{MC}$
in each point, and do not include the variance
of the fit parameters.} for different lattice sizes
and different $\beta$ values for $\beta \geq 2.30$.
In Figs.~1 and 2 our fits for $\beta_c$ (\ref{betac}) and
$(\sqrt{\sigma}a)$ (\ref{sigmamc}) (with the same parameter set
(\ref{min}) ) are compared with MC values of $\beta_c^{MC}$ and
$(\sqrt{\sigma}a)_{MC}$ from Tables I and II.

The standard criterion $\chi^{2} < \chi^{2}_{min} + 1 $ \cite{stat}
defines an confidence region of the model parameters which gives
a rather large variance of $g_f$,
$b$, $C_T$, $C_{\sigma}$, but with a very strong correlation
between them. In Fig.~3 we show the projection of this
confidence region to the $(g_f,b)$ plane. At the end point A,
$(g_f=0.288, b=0.800)$, we find $C_T=0.174$ and
$C_{\sigma}=0.253$.
Another end point B,
$(g_f=0.717, b=0.142)$, corresponds to
$C_T=8.90$ and $C_{\sigma}=12.98$.
Physical observables
stay almost unchanged under
large variance of the model parameters due to their strong correlations.
We find
\begin{equation}\label{ratio1}
T_{c}/\sqrt{\sigma}~\equiv~C_T/C_{\sigma}~=~0.687\pm 0.005~,
\end{equation}
where the error in Eq.~(\ref{ratio1}) caused
by the parameter variations is calculated from the variance
matrix of the parameters \cite{stat}.

From Eq.~(\ref{ansatz}) one can easily reconstruct the NP
beta function $\beta_f^{NP}(g)$
\begin{equation}\label{betanp}
\beta_f^{NP}~=~-~\frac{b_0^2 g^3}{b_0 - b_1 g^2 + 3 c_3 g^8}~.
\end{equation}
 In Fig.~4 we compare different
beta functions discussed in our paper: $\beta_f^{AF}(g)$ (\ref{betaaf}),
$\beta_f^{NP}(g)$ (\ref{betanp}) and $\beta_f^{FP}(g)$ (\ref{betafp}).
It is amazing that
$\beta_f^{NP}(g)$ with NP corrections (\ref{ansatz}) of
Ref.\cite{eng95} is very close to our straight line $\beta_f^{FP}(g)$
(\ref{betafp}) in a rather wide region of the coupling constant. Note
also that different numerical values of $T^{*}_{c}/\Lambda_{L}^{NP}$
and $T^{*}_{c}/\Lambda^{FP}_{L}$ are caused by a freedom in choosing
the form of expression (\ref{afp}) defining the arbitrary integration
constant of differential equation (\ref{beta}).

At the present moment there are no SU(2) MC data
for the critical temperature and the string tension
in the region $\beta > 2.85$.
In the framework of our FP scenario
we can predict the values of $\sqrt{\sigma}a$
and $\beta_c$ for the future MC calculations.
They are presented in Tables III and IV.
As is seen from Figs.~1 and 2
these numbers can be hardly distinguished
from those obtained with NP corrections (\ref{ansatz})
\cite{eng95} to the AF relation.
One needs very large values of $\beta$
to observe the difference.
The principal difference is of course exist:
$\beta_c \rightarrow \infty$ for
$N_{\tau}\rightarrow \infty$
in the approach of Ref.~\cite{eng95} and
$\beta_c \rightarrow 4/g_f^2~ =~const $
for $N_{\tau}\rightarrow \infty$ in the FP scenario.

It is clear that obtaining MC data for the critical temperature and
string tension in the $SU(2)$ gauge theory for
the region of $\beta$ essentially greater than 2.85
would require very large lattices and hardly possible
in the nearest future. It seems that these restrictions
are not so severe for the finite volume observables in
 the SU(2) gauge theory. In Ref.~\cite{finite}
the quantity $\bar{g}^2(L)$ defined as the response of the system in
hypercube $L\times L\times L\times L$ to a constant color-electric
background field was studied for $\beta =2.6 \div  3.7$.
To check the consistency of our FP scenario with these results
we reanalyzed some MC data obtained in Ref.~\cite{finite}.
Namely we use in our analysis the sets of
the bare couplings $\beta$ tuned to achieve constant values of
$\bar{g}^2(L_{j})$, ($j=0,2,4,6,8$) \footnote{We enumerate
$L_{j}$ with successive even numbers to retain the
notations of Ref. \cite{finite}.} for different lattice sizes
$N_{\sigma}=N_{\tau}\equiv N$.

From Eq.~ (\ref{afp}) it follows that the dependence
of $g$ on $N$ at constant $\bar{g}^2(L_j)$ is given by
\begin{equation}\label{gjfp}
g_{j}(N)~=~
\left( \frac{L_{j}\Lambda^{FP}}{N}  \right)^{b} + g_{f}~,
\end{equation}
where relation $a=L/N$ has been used.
In contrast to our previous consideration, the cutoff dependence (lattice
artifact) is not negligible in the present case.
Following to Ref.~\cite{finite} we assume that it is proportional
to $1/N$. Therefore to fit the data from the Table V we use
the following formula
\begin{equation}\label{betan}
\beta_{j}(N) = \frac{4}{(g_{j}(N))^{2}} - \frac{C_{j}}{N},
\end{equation}
where $C_{j}$ is a constant.

In our fit procedure
the FP beta function parameters $g_{f}$ and $b$ have been fixed at their
values (\ref{min}) found from the previous analysis, and
$L_{j}\Lambda^{FP}$ and $C_{j}$
for each column of Table V  are chosen to
minimize the following expression
\begin{equation}\label{chi2j}
\chi_{j}^{2}~=~
\sum_{N} \frac{[\beta_{j} - \beta_{j}^{MC}]^{2}}
{[\Delta \beta_{j}^{MC}]^{2}}.
\end{equation}
The results of the fit are shown in Table VI and in Fig.~5.

For a comparison we have made the similar fit assuming AF behavior of
$g_{j}(N)$, i.e. instead of Eq.~ (\ref{gjfp}) the value of
$g_{j}(N)$ was determined as a solution of the transcendental
equation
\begin{equation}\label{gjaf}
\frac{L_{j}\Lambda^{AF}}{N} = R\left( g_{j}^{2}(N) \right)
\end{equation}
with function $R$ defined by Eq.~(\ref{agaf}).
As is seen from Table VI, $\chi^{2}_{8}/\mbox{dof}$ is almost two
times smaller in the FP scenario then that in AF one. This value
correspond to the region $1.17 \le g \le 1.25$, where the difference
between AF and FP beta-functions becomes large (See Fig.~4). For the
rest of the data sets both approaches give nearly the same fit quality.
Still, the FP scenario provides a better agreement of the ratios
$L_{j}/L_{8}$ with those of Ref.\cite{finite} shown in the last
column of Table VI. Therefore, MC data of Ref.~\cite{finite}
are consistent with FP beta function behavior
(\ref{betafp},\ref{min}). In the region of $g\approx 1$
we do not observe the difference between AF and FP scenarios:
as seen from Fig.~4 $\beta_f^{AF}$ and $\beta_f^{FP}$
are close to each other in this region of $g$.
Hopefully additional lattice calculations in the spirit of
Ref.\cite{finite} at bare coupling $g<1$ would allow
one to find the true zero of the beta function $\beta_f(g)$.

We conclude that available MC lattice data in the
SU(2)-gluodynamics do not exclude the possibility of
the FP scenario (\ref{afp}). New lattice data are
necessary to prove (or disprove) the AF relation (\ref{agaf}).

\acknowledgements
The authors are thankful to  A. Bugrij, V.~Gusynin, O.~Mogilevslky
and B.~Svetitsky for fruitful discussions.

\newpage
\begin{table}\label{tab1}

\caption {MC data for critical couplings
$\beta_{c}^{MC}$ at different $N_{\tau}$ are taken from
Ref.~[1]. The values of $T_{c}/\Lambda_{L}^{AF}$
are calculated from Eq.~(\ref{Tcaf}) assumed the perturbative
AF relation (\ref{agaf}) (see also Ref.~[1]).
$T_{c}/\Lambda_{L}^{NP}$ are obtained from Eq.~(\ref{TcL}). Our results
for $T_{c}/\Lambda_{L}^{FP}$ followed from Eq.~(\ref{Tcfp}) in
the FP scenario are presented in the last column.}

\begin{tabular}{|c|c|c|c|c|}
$N_{\tau}$ & $\beta_c^{MC}$  & $T_c/\Lambda_L^{AF}$ &
$T_c/\Lambda_L^{NP}$ & $T_c/\Lambda_L^{FP}$ \\
\tableline
 2 & 1.8800(30)~ & 29.7 & 8.65(12)& 1.349(16) \\
 3 & 2.1768(30)~ & 41.4 &18.69(21)& 2.696(29) \\
 4 & 2.2988(6)~~ & 42.1 &21.44(05)& 3.084 (7) \\ \tableline
 5 & 2.3726(45)~ & 40.6 &21.95(33)& 3.167(48) \\
 6 & 2.4265(30)~ & 38.7 &21.81(22)& 3.156(32) \\
 8 & 2.5115(40)~ & 36.0 &21.44(27)& 3.124(41) \\
16 & 2.7395(100) & 32.0 &21.50(64)& 3.200(99) \\
\end{tabular}
\end{table}

\begin{table}\label{tab2}

\caption {MC data of
$(\sqrt{\sigma} a)_{MC}$
for different lattices
and coupling constants are taken from Ref.~[1].
The values $\sqrt{\sigma}/\Lambda_{L}^{AF}$ and
$\sqrt{\sigma}/\Lambda_{L}^{NP}$ are calculated from Eqs.(\ref{sigmaaf})
and (\ref{sigmaL}), respectively. Last column corresponds to
our results for
$\sqrt{\sigma}/\Lambda_{L}^{FP}$ (\ref{sqsifp}) in the FP scenario.}

\begin{tabular}{|c|c|c|c|c|c|c|}
$N_{\sigma}$ & $N_{\tau}$ & $\beta$  & $(\sqrt{\sigma} a)_{MC}$ &
$\sqrt{\sigma}/\Lambda_L^{AF}$   &
$\sqrt{\sigma}/\Lambda_L^{NP}$ &
$\sqrt{\sigma}/\Lambda_L^{FP}$  \\ \tableline
 8 & 10 & 2.20 & 0.4690(100) & 61.7(14) &28.56 (61) & 4.116(88)  \\ \tableline
10 & 10 & 2.30 & 0.3690(30)~ & 62.4(5)  &31.78 (26) & 4.574(38)  \\
16 & 16 & 2.40 & 0.2660(20)~ & 57.8(4)  &31.94 (25) & 4.615(35)  \\
32 & 32 & 2.50 & 0.1905(8)~~ & 53.3(2)  &31.51 (14) & 4.587(20)  \\
20 & 20 & 2.60 & 0.1360(40)~ & 49.0(14) &30.70 (91) & 4.509(133) \\
32 & 32 & 2.70 & 0.1015(10)~ & 47.1(5)  &31.03 (31) & 4.601(46)  \\
48 & 56 & 2.85 & 0.0630(30)~ & 42.8(21) &30.00(143) & 4.511(215) \\
\end{tabular}
\end{table}

\begin{table}\label{tab3}
\caption {Predictions for $(\sqrt{\sigma}a)_{MC}$
at different $\beta$ according to Eq.~(\ref{sigmamc}).
}
\begin{tabular}{|c|c|c|c|c|c|}
$\beta$ & 2.90 & 2.95 & 3.00 & 3.05 & 3.10 \\ \tableline
$\sqrt{\sigma}a$  & 0.0552(14) &
0.0476(15) & 0.0411(16) & 0.03555(17) & 0.0308(17)\\
\end{tabular}
\end{table}

\begin{table}\label{tab4}
\caption {Predictions for $\beta_c^{MC}$ at different $N_{\tau}$
according to Eq.~(\ref{betac}).
}
\begin{tabular}{|c|c|c|c|c|}
$N_{\tau}$ & 20 & 24 & 28 & 32 \\ \tableline
$\beta_c$ & 2.8077(49) & 2.8683(71) & 2.9201(93) & 2.9653(115)\\
\end{tabular}
\end{table}

\begin{table}\label{tab5}
\caption {The bare coupling $\beta$  at different lattice size $N$
for fixed $\bar{g}^2(L)$ [7]. The uncertainties of $\beta$ in
columns 2--5 were recalculated from errors of $\bar{g}^2(L)$, given
in Ref.~[7], using linear interpolation for the dependence
$\bar{g}^2(L)$ on $\beta$ at fixed $N$.}.
\begin{tabular}{|r|c|c|c|c|c|}
$N$ & \multicolumn{5}{c|}{$\beta$}\\ \tableline
    & $\bar{g}^2(L_{0})=2.037$
    & $\bar{g}^2(L_{2})=2.380$
    & $\bar{g}^2(L_{4})=2.840$
    & $\bar{g}^2(L_{6})=3.550$
    & $\bar{g}^2(L_{8})=4.765$  \\ \tableline
 5 & 3.4564(25) & 3.1898(22) & 2.9568(19) & 2.7124(34) &            \\
 6 & 3.5408(40) & 3.2751(35) & 3.0379(31) & 2.7938(48) & 2.5752(28) \\
 7 & 3.6045(42) & 3.3428(36) & 3.0961(33) & 2.8598(50) & 2.6376(20) \\
 8 & 3.6566(47) & 3.4009(42) & 3.1564(38) & 2.9115(55) & 2.6957(21) \\
10 & 3.7425(59) & 3.5000(61) & 3.2433(53) & 3.0071(77) & 2.7824(22) \\
12 &            &            &            &            & 2.8485(32) \\
14 &            &            &            &            & 2.9102(62) \\
\end{tabular}
\end{table}

\begin{table}\label{tab6}
\caption {The results of fitting the data from Table V with
Eq.~(\ref{betan}) assuming FP (\ref{gjfp}) and
AF (\ref{gjaf}) scenario. The ratios $L_{j}/L_{8}$ are compared with
those calculated in Ref.~[7]. The number of degrees of freedom (dof),
the difference between the number of points and the number of fit
parameters, equals 3 for $j=0,2,4,6$ and 4 for $j=8$.}
\begin{tabular}{|r|l|l|l|l|l|l|l|l|l|}
 & \multicolumn{4}{c|}{FP}
 & \multicolumn{4}{c|}{AF}& Ref.~\cite{finite}\\ \tableline
$j$ & $\chi_{j}^{2}/\mbox{dof}$ & $C_{j}$ &
$L_{j}\Lambda_{L}^{FP}\times 10$ & $L_{j}/L_{8}$ &
  $\chi_{j}^{2}/\mbox{dof}$ &
$C_{j}$ & $L_{j}\Lambda_{L}^{AF}\times 10^{2}$ &  $L_{j}/L_{8}$ &
  $L_{j}/L_{8}$ \\ \tableline
0 & 1.25  & 0.315 & 0.102 &  0.072  &
                   1.08  &  0.215  & 0.139 & 0.080 & 0.070(8)\\
2 & 0.09  & 0.559 & 0.187 &  0.131  &
                   0.09  &  0.378  & 0.254 & 0.146 & 0.124(13)\\
4 & 0.75  & 0.422 & 0.387  &  0.272  &
                   0.81  &  0.177  & 0.512  & 0.293 & 0.249(19)\\
6 & 0.14  & 0.585 & 0.734  &  0.516  &
                   0.18  &  0.220  & 0.933  & 0.534 & 0.500(23)\\
8 & 0.89  & 0.589 & 1.442  &  1.000  &
                   1.67  &  0.032  & 1.742  & 1.000 & 1.000
\end{tabular}
\end{table}

\begin{figure}[t]\label{fig1}
\begin{center}
\vfill
\leavevmode
\epsfysize=15cm \epsfbox[80 390 440 720]{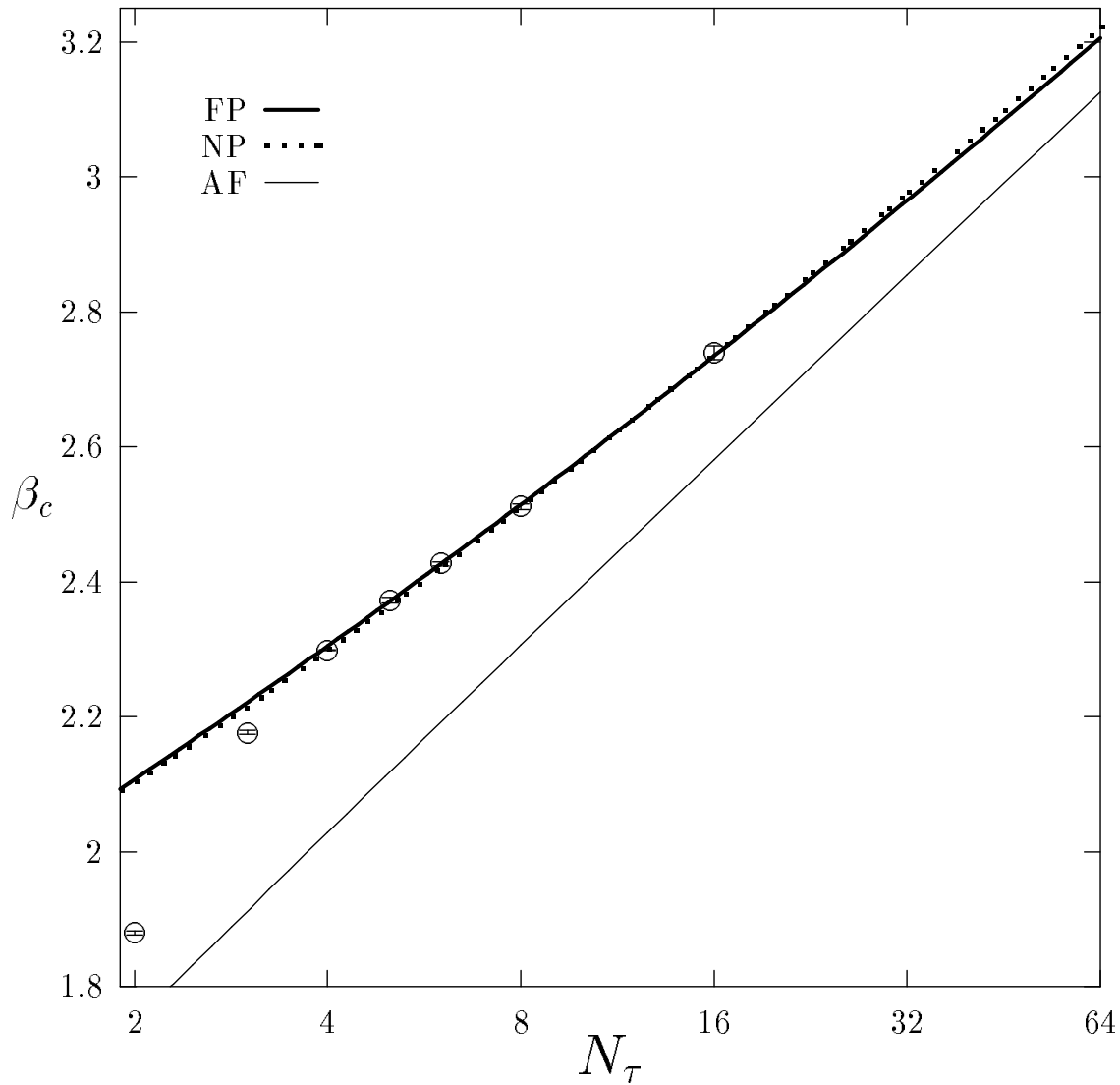}
\vfill
\caption{Circles with errorbars represent the
MC data listed in Table I.
The bold solid line corresponds to Eq.~(\ref{betac})
of the FP scenario.
The dotted and thin solid
lines show $\beta_c=4/g_c^2$ found
by solving the equations
$T_c^*/\Lambda_L^{NP}=[N_{\tau}\lambda(g^2_c)R(g_c^2)]^{-1}$
and
$T_c^*/\Lambda_L^{NP}=[N_{\tau}R(g_c^2)]^{-1}$, respectively,
with $T_c^*/\Lambda_L^{NP}=21.45$.
}
\end{center}
\end{figure}

\begin{figure}[t]\label{fig2}
\mbox{}
\begin{center}
\vfill
\leavevmode
\epsfysize=15cm \epsfbox[80 390 440 720]{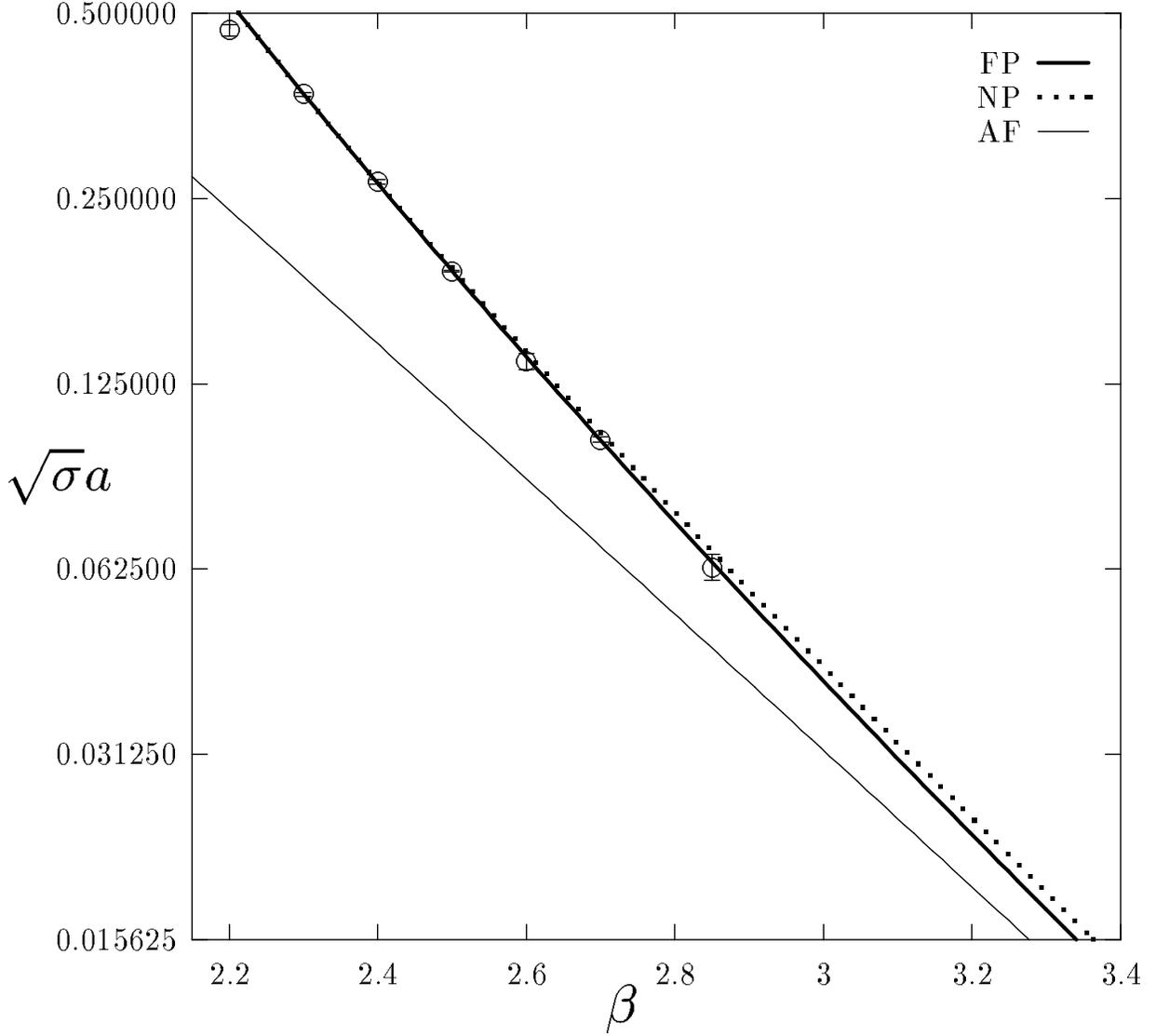}
\vfill
\caption{Circles with errorbars represent the
MC data listed in Table II.
The bold solid line corresponds to Eq.~(\ref{sigmamc})
of the FP scenario.
The dotted and thin solid
lines are calculated
as $\lambda(g^2)R(g^2)\sqrt{\sigma^*}/\Lambda_L^{NP}$
and
$R(g^2)\sqrt{\sigma^*}/\Lambda_L^{NP}$, respectively,
with
$\sqrt{\sigma ^*}/\Lambda_L^{NP}=31.56$.}
\end{center}
\end{figure}

\begin{figure}[t]\label{fig3}
\mbox{}
\begin{center}
\vfill
\leavevmode
\epsfysize=15cm \epsfbox[80 390 440 720]{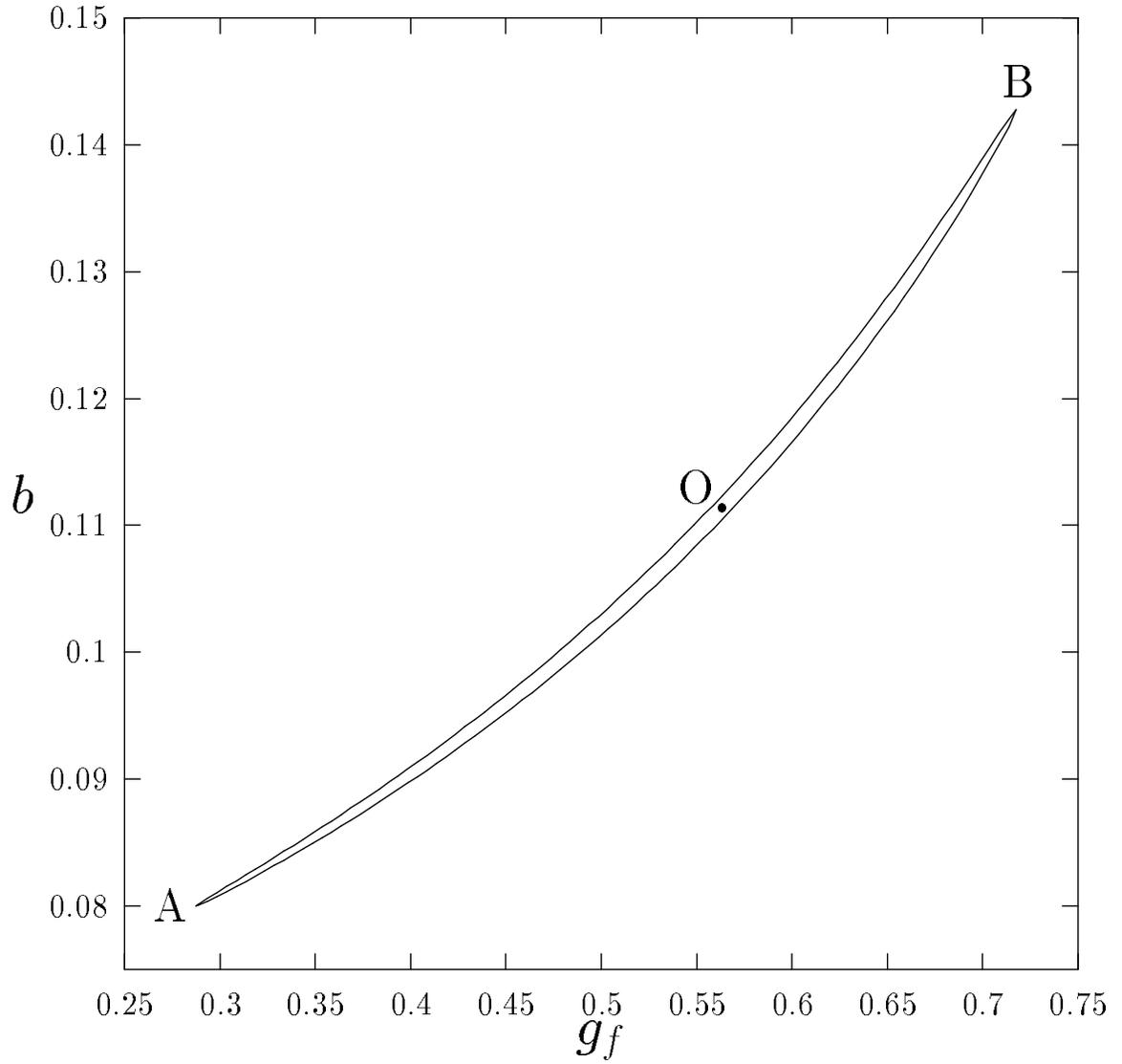}
\vfill
\caption{Projection of the confidence region of the parameters to the
$(g_f,b)$ plane. The point $O$ corresponds to the best fit (\ref{min}).
}
\end{center}
\end{figure}

\begin{figure}[t]\label{fig4}
\mbox{}
\begin{center}
\vfill
\leavevmode
\epsfysize=15cm \epsfbox[80 390 440 720]{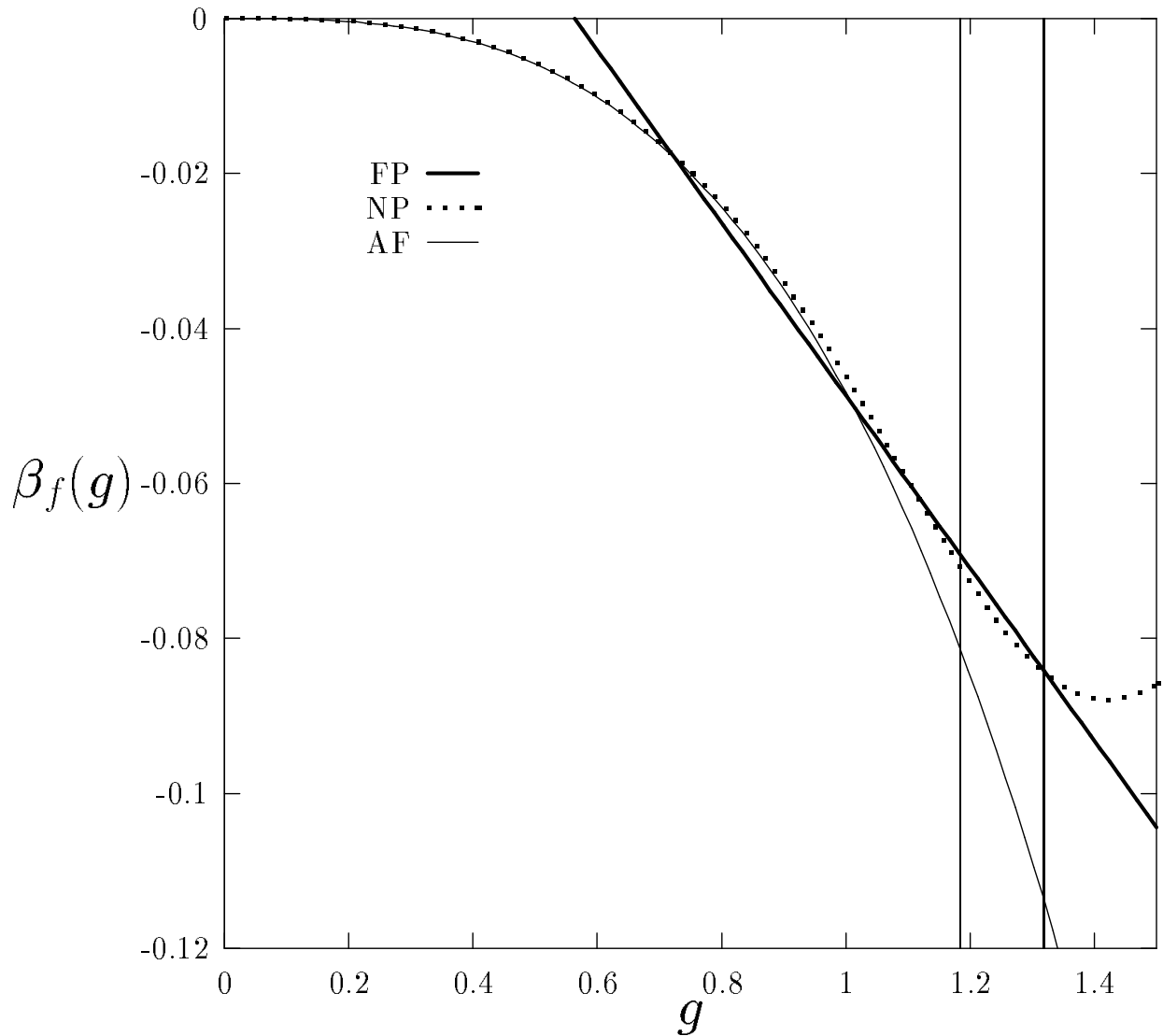}
\vfill
\caption{The behavior of $\beta_{f}^{FP}(g)$ (\ref{betafp})
(bold solid line),
$\beta_{f}^{NP}(g)$ (\ref{betanp})
(dotted line) and $\beta_{f}^{AF}(g)$ (\ref{betaaf}) (thin solid
line).  Two vertical lines show the region of the coupling
constant ($\beta=2.3\div2.8$) of MC data used in the fitting
procedures.}
\end{center}
\end{figure}

\begin{figure}[t]\label{fig5}
\mbox{}
\begin{center}
\vfill
\leavevmode
\epsfysize=15cm \epsfbox[105 378 440 725]{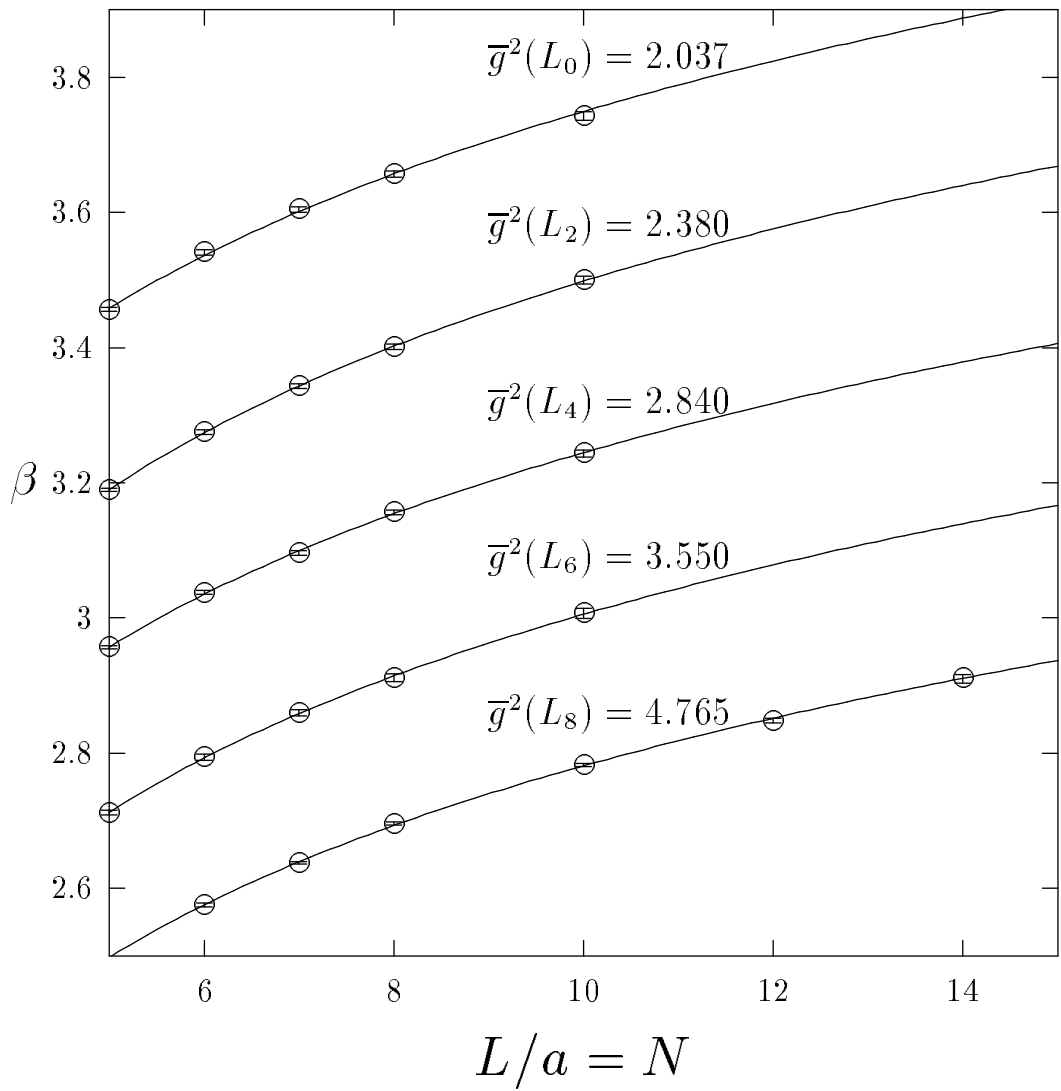}
\vfill
\caption{The results of fitting the data from Table V with
the formulas (\ref{gjfp}) and (\ref{betan}).}
\end{center}
\end{figure}

\end{document}